\documentclass[12pt,a4paper]{article}
\usepackage{amssymb,amsmath}
\usepackage{graphicx}

\def\openone{\leavevmode\hbox{\small1\kern-3.3pt\normalsize1}}

\def\b{\begin{eqnarray}}
\def\e{\end{eqnarray}}
\def\n{\noindent}

\newtheorem{theorem}{Theorem}[section]

\begin{document}

\begin{center}

{\Large \bf Generalised Fourier Transform and Perturbations \\[10pt]
to Soliton Equations}

\bigskip

{\large \bf  Georgi G. Grahovski$^{\dag}$\footnote{E-mail: {\tt grah@inrne.bas.bg}}\footnote{On leave from Institute for Nuclear Research and Nuclear Energy, Bulgarian Academy of Sciences, 72 Tsarigrasko chaussee, 1784 Sofia, Bulgaria} and Rossen I. Ivanov$^{\ddag}$\footnote{E-mail: {\tt rivanov@dit.ie}} }

\end{center}
\medskip

\n {\it $^{\dag}$ School of Electronic Engineering, Dublin City
University, Glasnevin, Dublin 9, Ireland}

\n {\it $^{\ddag}$ School of Mathematical Sciences, Dublin Institute of Technology, Kevin Street, Dublin 8, Ireland}

\bigskip
\begin{abstract}
\noindent A brief survey of the theory of soliton perturbations is
presented. The focus is on the usefulness of the so-called
Generalised Fourier Transform (GFT). This is a method that
involves expansions over the complete basis of ``squared
solutions'' of the spectral problem, associated to the soliton
equation. The Inverse Scattering Transform for the corresponding
hierarchy of soliton equations can be viewed as a GFT where the
expansions of the solutions have generalised Fourier coefficients
given by the scattering data.

The GFT provides a natural setting for the analysis of small
perturbations to an integrable equation: starting from a purely
soliton solution one can 'modify' the soliton parameters such as
to incorporate the changes caused by the perturbation.

As illustrative examples the perturbed equations of the KdV
hierarchy, in particular the Ostrovsky equation, followed by the
perturbation theory for the Camassa-Holm hierarchy are presented.

\medskip


\medskip

\end{abstract}


\section{Introduction}\label{sec:1}

Integrable equations are widely used as model equations in various
problems. The integrability concept originates from the fact that
these equations are in some sense exactly solvable, e.g. by the
inverse scattering method (ISM), and exhibit global regular
solutions. This feature is very important for applications, where
in general analytical results (first integrals, particular
solutions) are preferable to numerical computations, which are not
only long and costly, but also intrinsically subject to numerical
error. In a hydrodynamic context, even though water waves are
expected to be unstable in general, they do exhibit certain
stability properties in physical regimes where integrable model
equations are accurate approximations for the evolution of the
free surface water wave cf. \cite{AL08,CL09}.

There are situations however where the model equation is not
integrable, but is somehow close to an integrable equation, i.e.
can be considered as a perturbation of an integrable equation. In
such case it is still possible to obtain approximate analytical
solutions. There are several approaches treating the perturbations
of integrable equations. One possibility is to consider expanding
the solutions of the perturbed nonlinear equation around the
corresponding unperturbed solution and to determine the
corrections due to perturbations. In other words, one represents
the solutions $\tilde{u}(x,t)$ in the form: \b \tilde{u}(x,t) =
u(x,t)+\Delta u(x,t), \nonumber \e where $u(x,t)$ is the solution
of the corresponding unperturbed nonlinear evolutionary equation
and $\Delta u(x,t)$ is a perturbation. The strength of the
perturbation is measured by a parameter $\epsilon$, $\Delta
u(x,t)=\mathcal{O}(\epsilon)$. By {\it small (weak) perturbation}
one means $0<\epsilon \ll 1$. Such perturbations can be studied
{\it directly} in the configuration (coordinate) space, while the
effect of the perturbations on the corresponding scattering data
can be studied in the {\it spectral space} (usually the complex
plane of the spectral parameter) of the associated spectral
problem.

For a {\it direct} study of soliton perturbations, one can use the
multi-scale expansion method \cite{degasp,dms1997}, introducing multiple scales,
i.e. transforming the independent time variable $t$ into several
variables $t_n, (n=0,1,2,\dots)$ by
\[
t_n=\epsilon^n t, \qquad n=0,1,2,\dots,
\]
where each $t_n$is an order of $\epsilon$ smaller than the
previous time $t_{n-1}$. Then, the time-derivative are replaced by
the expansion (the so-called ``derivative expansion'') with
respect to the multiple scales:
\[
\partial_{t}=\sum_{n=0}^{\infty}\epsilon^n\partial_{t_n}.
\]
The dependent variable is expanded in an asymptotic series
\[
u(x,t)=\sum_{n=0}^{\infty}\epsilon^nu_{n}(x,t).
\]
These expressions are substituted back into the equation, giving a
sequence of equations for $u_n(x,t)$, corresponding to each order
of $\epsilon$ (each time scale $t_n=\epsilon t$). Solving the
system of equations for $u_n(x,t)$, one has to ensure that there
are no singularities in the solutions (i.e. that the solutions do
not blow up in time, etc.). This may lead to some additional
conditions on the functions $u_n(x,t)$ (or on the parameters in
them), known as {\it secular conditions}.

Several authors had used various versions of the direct approach
in the study of soliton perturbations: D. J. Kaup \cite{kaup1983}
had used a similar approach for the perturbed sine-Gordon
equation. Keener and McLaughlin \cite{knm1977a} had proposed a
direct approach by obtaining the appropriate Green functions for
the nonlinear Schrodinger and sine-Gordon equations. For a
comprehensive review of the direct perturbation theory see e.g.
\cite{degasp,He90a} and the references therein.

In the spectral space, the study of the soliton perturbations is
based on the perturbations of the scattering data, associated to
the spectral problem. Such methods are used by a number of
authors, for studying perturbations of various nonlinear
evolutionary equations, like the sin-Gordon equation
\cite{kaup1977}, the nonlinear Schr\"odinger equation
\cite{knm1977b,GI92,g1994} and of course, the KdV equation which
is discussed in details in the following sections. The method is
based on expanding the 'potential' (i.e. the dependent variable)
$u(x,t)$ of the associated spectral problem over the complete set
of ``squared solutions'', which are eigenfunctions of the
corresponding recursion operator.

The squared eigenfunctions of the spectral problem associated to
an integrable equation represent a complete basis of functions,
which helps to describe the Inverse Scattering Transform for the
corresponding hierarchy as a Generalised Fourier transform (GFT).
The Fourier modes for the GFT are the Scattering data. Thus all
the fundamental properties of an integrable equation such as the
integrals of motion, the description of the equations of the whole
hierarchy and their Hamiltonian structures can be naturally
expressed making use of the completeness relation for the squared
eigenfunctions and the properties of the recursion operator.

The GFT also provides a natural setting for the analysis of small
perturbations to an integrable equation.  The leading idea is that
starting from a purely soliton solution of a certain integrable
equation one can 'modify' the soliton parameters such as to
incorporate the changes caused by the perturbation. There is a
contribution to the equations for the scattering data that comes
from the GFT-expansion of the perturbation.

In this review article we illustrate these ideas with several
examples. Firstly we consider the equations of the KdV hierarchy
and the KdV perturbed version -- the Ostrovsky equation. Then we
present the perturbation theory for the Camassa-Holm hierarchy.

\section{Basic facts for the inverse scattering method for the KdV hierarchy}\label{sec:2}

\subsection{Direct scattering transform and scattering data}\label{ssec:2.1}

The spectral problem for the equations of the KdV hierarchy is
\cite{ZMNP,IKK94} \b \label{SPKdV} -\Psi_{xx}+u(x)\Psi=k^2\Psi, \e

\n in which the real-valued potential $u(x)$ is taken for
simplicity to be a function of Schwartz-type: $u(x)\in
\mathcal{S}(\mathbb{R})$, $k \in \mathbb{C}$ is spectral
parameter. The continuous spectrum under these conditions
corresponds to real $k$. The discrete spectrum consists of
finitely many points $k_{n}=i\kappa _{n}$, $n=1,\ldots,N$ where
$\kappa_{n}$ is real.

The Jost solutions for (\ref{SPKdV}) are as follows: $f^+(x,k)$
and $\bar{f}^+(x,\bar{k})$  are fixed by their asymptotic when
$x\rightarrow\infty$ for all real $k\neq 0$ \cite{ZMNP}: \b
\label{eq6} \lim_{x\to\infty }e^{-ikx} f^+(x,k)= 1, \e
$f^-(x,k)$ and$\bar{f}^-(x,\bar{k})$ fixed by their asymptotic
when $x\rightarrow -\infty$ for all real $k\neq 0$: \b
\label{eq6'} \lim_{x\to -\infty }e^{ikx} f^-(x,k)= 1, \e

\n Since $u(x)$ is real then
\begin{equation}\label{eq:inv}
 \bar{f}^+(x,\bar{k}) = f^+(x,-k), \qquad \mbox{and} \qquad
 \bar{f}^-(x,\bar{k}) = f^-(x,-k).
\end{equation}

\n In particular, for real $k\neq 0$ we have:

\b \label{eq5aa} \bar{f}^{\pm}(x,k)=f^{\pm}(x, -k), \e

\n and the vectors of the two bases are related
\footnote{According to the notations used in \cite{ZMNP}
$f^+(x,k)\equiv \bar{\psi}(x,\bar{k})$, $f^-(x,k)\equiv
\varphi(x,k)$.}:

\b \label{eq8} f^{-}(x,k)=a(k)f^+(x,-k)+b(k)f^+(x,k), \qquad
\mathrm{Im}\phantom{*} k=0. \e
The coefficient $a(k)$ allows analytic extension in the upper half
of the complex $k$-plane  \cite{ZMNP} and
\begin{equation}\label{eq:inv1}
\bar{a}(\bar{k}) = a(-k), \qquad \bar{b}(\bar{k}) = b(-k), \qquad
|a(k)|^{2}-|b(k)|^{2}=1.
\end{equation}
The quantities $\mathcal{R}^{\pm}(k)=b(\pm k)/a(k)$ are known
as reflection coefficients (to the right with superscript ($+$) and
to the left with superscript ($-$) respectively). It is sufficient
to know $\mathcal{R}^{\pm}(k)$ only on the half line $k>0$, since
from (\ref{eq:inv1}):
$\mathcal{R}^{\pm}(-k)=\bar{\mathcal{R}}^{\pm}(k)$ and also
(\ref{eq:inv1}) \b \label{eq13a}
|a(k)|^{2}=(1-|\mathcal{R}^{\pm}(k)|^{2})^{-1}, \e

\n Furthermore $\mathcal{R}^{\pm}(k)$ uniquely determines $a(k)$
\cite{ZMNP}.  At the points $\kappa_n$ of the discrete spectrum,
$a(k)$ has simple zeroes i.e.:
\begin{equation}\label{eq:a-n}
    a(k) = (k-i\kappa_n)\dot{a}_n +\frac{1}{2} (k-i\kappa_n)^2\ddot{a}_n
    + \cdots,
\end{equation}

\n The dot stands for a derivative with respect to $k$ and
$\dot{a}_n\equiv \dot{a}(i\kappa_n)$, $\ddot{a}_n\equiv
\ddot{a}(i\kappa_n)$, etc. The following dispersion relation
holds:  \b a(k)=\exp \left( -\frac{1}{2\pi i}
\int_{-\infty}^{\infty}\frac{\ln(1-|\mathcal{R}^{\pm}(k')|^2)}{k'-k}\mathrm{d}k'
\right)\prod_{j=1}^{N}\frac{k-i\kappa_j}{k+i\kappa_j}.\label{dr}\e

\n At the points of the discrete spectrum $f^-$ and $f^+$ are
linearly dependent: \b \label{eq200}
f^-(x,i\kappa_n)=b_nf^+(x,i\kappa_n).\e

\n  In other words, the discrete spectrum is simple, there is only
one (real) linearly independent eigenfunction, corresponding to
each eigenvalue $i\kappa_n$, say \b \label{eq201}f_n^-(x)\equiv
f^-(x,i\kappa_n)\e

\n From (\ref{eq201}) and (\ref{eq6}), (\ref{eq6'}) it follows
that $f_n^-(x)$ falls off exponentially for $x\to\pm\infty$, which
allows one to show that $f_n(x)$ is square integrable. Moreover,
for compactly supported potentials $u(x)$ (cf. (\ref{eq200}) and
(\ref{eq8})) \b \label{eq202} b_n= b(i\kappa_n), \qquad
b(-i\kappa_n)=-\frac{1}{b_n} .\e

\n One can argue \cite{ZMNP}, that the results from this case can
be extended to Schwarz-class potentials by an appropriate limiting
procedure.

\n The asymptotic of $f_n^-$, according to (\ref{eq5aa}),
(\ref{eq6}), (\ref{eq200}) is \b \label{eq203}
f_n^-(x)&=&e^{\kappa_n x}+o(e^{\kappa_n x}), \qquad x\rightarrow
-\infty;
 \\\label{eq204}
f_n^-(x)&=& b_n e^{-\kappa_n x}+o(e^{-\kappa_n x}), \qquad
x\rightarrow \infty. \e

\n The sign of $b_n$ obviously depends on the number of the zeroes
of $f_n^-$. Suppose that
$0<\kappa_{1}<\kappa_{2}<\ldots<\kappa_{N}$. Then from the
oscillation theorem for the Sturm-Liouville problem \cite{B},
$f_n^-$ has exactly $n-1$ zeroes. Therefore \b \label{eq205} b_n=
(-1)^{n-1}|b_n|.\e

\n The sets \b \label{eq206} \mathcal{S^{\pm}}\equiv\{
\mathcal{R}^{\pm}(k)\quad (k>0),\quad \kappa_n,\quad
R_n^{\pm}\equiv\frac{b_n^{\pm1}}{i\dot{a}_n},\quad n=1,\ldots N\}
\e

\n are called scattering data. Clearly, due to (\ref{dr}) each set
-- $\mathcal{S^{+}}$ or $\mathcal{S^{-}}$ of scattering data
uniquely determines the other one and also the potential $u(x)$
\cite{ZMNP,IKK94,ZF71}.

\subsection{Generalised Fourier Transform}\label{ssec:2.2}

The recursion operator for the KdV hiererchy is \b
L_{\pm}=-\frac{1}{4}\partial^2+u(x)-\frac{1}{2}\int_{\pm
\infty}^{x}\mathrm{d}\tilde{x} u'(\tilde{x})\cdot. \label{recurs} \e

\n The eigenfunctions of the recursion operator are the squared
eigenfunctions of the spectral problem:
\begin{equation}\label{eq23} F^{\pm}(x,k)\equiv (f^{\pm}(x,k))^2,
\qquad F_n^{\pm}(x)\equiv F(x,i\kappa_n),
\end{equation}

\n where $F_n^{\pm}(x)$ are related to the discrete spectrum
$k_n=i\kappa_n$. Using (\ref{SPKdV}) and the asymptotics
(\ref{eq6}), (\ref{eq6'}) one can show that
\begin{equation}\label{eq23a} L_{\pm}F^{\pm}(x,k)=k^2F^{\pm}(x,k)\qquad L_{\pm}F^{\pm}_n(x)=k_n^2F^{\pm}_n(x).
\end{equation}
The Generalised Fourier expansion can be formulated as follows:

\begin{theorem} \label{GFE} Suppose that $f^+$ and $f^-$ are not linearly dependent at $x=0$. For each function $g(x)\in
\mathcal{S}(\mathbb{R})$ the following expansion formulas hold: \b
g(x)=\pm \frac{1}{2\pi
i}\int_{-\infty}^{\infty}\tilde{g}^{\pm}(k)F^{\pm}_{x}(x,k)\mathrm{d}k\mp\sum_{j=1}^{N}\left(g^{\pm}_{1,j}\dot{F}^{\pm}_{j,x}(x)+
g^{\pm}_{2,j}F^{\pm}_{j,x}(x) \right), \nonumber \e

\n where $\dot{F}^{\pm}_{j}(x)\equiv [\frac{\partial}{\partial k}
F^{\pm} (x,k)]_{k=k_j}$ and the Fourier coefficients are

\b \tilde{g}^{\pm}(k)&=&\frac{1}{k a^2(k)}\left(g,F^{\mp} \right),
\quad \mathrm{where} \quad \left(g,F\right)\equiv \int_{-\infty}^{\infty} g(x)F(x) \mathrm{d}x, \nonumber \\
g^{\pm}_{1,j}&=&\frac{1}{k_j \dot{a}_j^2}\left(g,F^{\mp}_j
\right),\nonumber \\
g^{\pm}_{2,j}&=&\frac{1}{k_j
\dot{a}_j^2}\left[\left(g,\dot{F}^{\mp}_j\right)-\left(\frac{1}{k_j}+\frac{\ddot{a}_j}{\dot{a}_j}\right)\left(g,F^{\mp}_j\right)
\right].\nonumber\e

\end{theorem}

{\bf Proof:}
The details of the derivation can be found e.g. in
\cite{E81,IKK94}.
$\blacksquare$

In particular one can expand the potential $u(x)$, the
coefficients are given through the scattering data
\cite{E81,IKK94}: \b u(x)=\pm \frac{2}{\pi
i}\int_{-\infty}^{\infty}k\mathcal{R}^{\pm}(k)F^{\pm}(x,k)\mathrm{d}k+\sum_{j=1}^{N}4ik_jR_j^{\pm}F^{\pm}_{j}(x).
\label{Exp u} \e

\n The variation  $\delta u(x)$ under the assumption that the
number of the discrete eigenvalues is conserved is \b \delta
u(x)=- \frac{1}{\pi }\int_{-\infty}^{\infty}\delta
\mathcal{R}^{\pm}(k)F^{\pm}_x(x,k)\mathrm{d}k\pm 2
\sum_{j=1}^{N}\left[R_j^{\pm}\delta
k_j\dot{F}^{\pm}_{j,x}(x)+\delta R_j^{\pm}F_{j,x}^{\pm}\right].
\label{Exp delta u}\e An important subclass of variations are due
to the time-evolution of $u$, i.e. effectively we consider a
one-parametric family of spectral problems, allowing a dependence
of an additional parameter $t$ (time). Then $\delta u(x,t)=u_t
\delta t + Q((\delta t)^2)$, etc. The equations of the KdV
hierarchy can be written as \b u_t+\partial_x
\Omega(L_{\pm})u(x,t)=0, \label{KdVH} \e where $\Omega(k^2)$ is a
rational function specifying the dispersion law of the equation.
The substitution of (\ref{Exp delta u}) and (\ref{Exp u}) in
(\ref{KdVH}) due to (\ref{eq23a}) gives a system of trivial linear
ordinary differential equations for the scattering data: \b
\mathcal{R}_t^{\pm}\pm
2ik\Omega(k^2)\mathcal{R}^{\pm}&=&0,\label{ScatData 1} \\
R_{j,t}^{\pm}\pm
2ik_j\Omega(k_j^2)R_j^{\pm}&=&0, \label{ScatData 2} \\
k_{j,t}&=&0.\label{ScatData 3} \e The KdV equation \b
\label{eq:KdV}u_t-6uu_x+u_{xxx}=0 \e can be obtained for
$\Omega(k^2)=-4k^2$.

Once the scattering data are determined from (\ref{ScatData 1}) --
(\ref{ScatData 3}) one can recover the solution from (\ref{Exp
delta u}). Thus the Inverse Scattering Transform can be viewed as
a GFT.

\section{Perturbations of the equations of the KdV hierarchy}\label{sec:3}

Let us now consider a perturbed equation from the KdV hierarchy,
i.e. an equation of the form \b u_t+\partial_x
\Omega(L_{\pm})u(x,t)=P[u], \label{PKdVH} \e

\n where $P[u]$ is a small perturbation, which is also assumed in
the Schwartz-class. The perturbed equation is, in general,
non-integrable. One can expand $P[u]$ according to Theorem
\ref{GFE} and $u_t$ and $u$ according to (\ref{Exp delta u}) and
(\ref{Exp u}). Their substitution in (\ref{PKdVH}) now apparently
leads to a modification of (\ref{ScatData 1}) -- (\ref{ScatData
3}) as follows:\b \mathcal{R}_t^{\pm}\pm
2ik\Omega(k^2)\mathcal{R}^{\pm}&=&\pm \frac{(P,F^{\mp})}{2ka^2(k)},\label{PScatData 1} \\
R_{j,t}^{\pm}\pm
2ik_j\Omega(k_j^2)R_j^{\pm}&=&-\frac{1}{2k_j\dot{a}_j^2}\left[(P,\dot{F}_j^{\mp})-
\left(\frac{1}{k_j}+\frac{\ddot{a}_j}{\dot{a}_j}\right)\left(P,F^{\mp}_j\right)\right], \label{PScatData 2} \\
k_{j,t}&=&-\frac{(P,F_j^{\mp})}{2k_j\dot{a}_j^2R_j^{\pm}}.\label{PScatData
3} \e

\n Note that due to (\ref{PScatData 3}) as a result of the
perturbation the discrete eigenvalues are time-dependent. Another
feature is the contribution from the continuous spectrum: even if
one starts with a pure soliton solution of the unperturbed
equation ($\mathcal{R}^{\pm}(k,0)=0$) then, in general
$\mathcal{R}^{\pm}(k,t)\ne 0$ due to (\ref{PScatData 1}).

For practical applications it is easier to work with an equation
for $b_n$ instead of (\ref{PScatData 2}). Such an equation can be
obtained as follows. We notice that \b R_{n,t}^+&=&\left(
\frac{b_n}{i\dot{a}_n}\right)_t=\frac{1}{i\dot{a}_n}b_{n,t}+b_n
\left(\frac{1}{i\dot{a}_n}\right)_t, \nonumber \\
R_{n,t}^-&=&\left(
\frac{1}{ib_n\dot{a}_n}\right)_t=-\frac{1}{i\dot{a}_nb_n^2}b_{n,t}+\frac{1}{b_n}
\left(\frac{1}{i\dot{a}_n}\right)_t, \nonumber \e

\n thus $b_{n,t}=\frac{i\dot{a}_n}{2}(R_{n,t}^+-b_n^2R_{n,t}^-)$.
Then using (\ref{PScatData 2}) and the fact that
$F_n^-=b_n^2F_n^+$, cf. (\ref{eq200}) we have

\b b_{n,t}+2ik_n\Omega(k_n^2)b_n=\frac{i}{4k_n \dot{a}_n} \left(P,
b_n^2\dot{F}_n^+ - \dot{F}_n^-\right). \label{b_n} \e
As an example let us consider the adiabatic perturbation of the
one-soliton solution of the KdV equation. The one-soliton solution
is \b u_s(x,t)=-2\kappa_1^2 \mathrm{sech} ^2 z, \qquad
z=\kappa_1(x-\xi), \qquad \xi=4\kappa_1^2t+\xi_0. \label{1s} \e
The eigenfunctions are

\b f^{\pm}(x,k)=\frac{e^{\pm ikx}(k\pm i\kappa_1 \tanh z)
}{k+i\kappa_1}, \quad a(k)=\frac{k-i\kappa_1}{k+i\kappa_1}, \quad
b_1=e^{2\kappa_1 \xi} .\e

\n The perturbed solution is \b u(x,t)=-2\kappa_1^2 \mathrm{sech}
^2 z + u_r(x,t), \qquad z=\kappa_1(t)[x-\xi(t)]. \label{p1s}\e

\n Here $u_r(x,t)$ is the contribution from the continuous
spectrum (radiation).

\n From (\ref{PScatData 3}) we have \b
\kappa_{1,t}=-\frac{1}{4\kappa_1}\int_{-\infty}^{\infty}
P[u_s(z)]\mathrm{sech}^2 z \mathrm{d} z. \label{kapa t} \e Writing
$b_1=e^{2\kappa_1(t)\xi(t)}$ and using (\ref{kapa t}) and
(\ref{b_n}) we obtain \b
\xi_t=4\kappa_1^2-\frac{1}{4\kappa_1^3}\int_{-\infty}^{\infty}
P[u_s(z)]\left(z+\frac{1}{2}\sinh 2z\right)\mathrm{sech}^2 z
\mathrm{d} z. \label{xi t} \e \n For the reflection coefficient
(\ref{PScatData 1}) gives \b \mathcal{R}_t^{+}-
8ik^3\mathcal{R}^{+}&=&
\frac{ie^{-2ik\xi}}{2k\kappa_1}\int_{-\infty}^{\infty}
P[u_s(z)]e^{-2ikz/\kappa_1}(k-i\kappa_1 \tanh z)^2 \mathrm{d} z.
\label{PReflCoeff} \e \n then according to \cite{KM77} using
approximations in Gelfand-Levitan-Marchenko equation one can
obtain \b u_r(x,t)=
\frac{\kappa_1}{\pi}\frac{\mathrm{d}}{\mathrm{d}
z}\int_{-\infty}^{\infty}\mathcal{R}^{+}(k)e^{2ik\xi+2ikz/\kappa_1}\left(\frac{k+i\kappa_1
\tanh z}{k+i\kappa_1}\right)^2 \mathrm{d}k. \e Alternatively, from
(\ref{Exp u}) it follows that \b u_r(x,t)&=&\frac{2}{\pi
i}\int_{-\infty}^{\infty}k\mathcal{R}^{+}(k)F^+(x,k)\mathrm{d}k
\nonumber \\ &=& \frac{2}{\pi
i}\int_{-\infty}^{\infty}k\mathcal{R}^{+}(k)e^{2ik\xi+2ikz/\kappa_1}\left(\frac{k+i\kappa_1
\tanh z}{k+i\kappa_1}\right)^2 \mathrm{d}k. \e Both formulae give
an approximation of the radiation component since the $z$-
derivative of $\tanh z$ can be neglected \cite{L80}.

The perturbation results for the Zakharov-Shabat (ZS) type
spectral problems have been obtained firstly in \cite{K76} and for
KdV in \cite{KM77}. As it has been explained, the perturbation
theory is based on the completeness relations for the squared
eigenfunctions. For the Sturm-Liouville spectral problem such
relations apparently have been studied as early as in 1946
\cite{B46} and then by other authors, e.g. \cite{KKh81,IKK94}. The
completeness relation for the eigenfunctions of the ZS spectral
problem is derived in \cite{K76a} and generalisations are studied
further in \cite{G86,GI92,GeHr1,GY94,TV08}, see also
\cite{gvy2008}.

{\bf Example:} {\it Ostrovsky equation.} This equation has the
form \cite{O78}:

\begin{equation}\label{ostr}
 u_{t}+ u_{xxx}-6uu_x=\gamma \partial^{-1}u,
\end{equation}

\n where $\partial^{-1}$ is an operator such that
$(\partial^{-1}u)_x\equiv u$, in general not uniquely determined.
The Ostrovsky equation can be viewed as a time-dependent nonlocal
perturbation of the KdV equation (\ref{eq:KdV}). Here $\gamma$  is
a constant parameter. The equation is often called the
Rotation-Modified Korteweg-de Vries equation. It describes gravity
waves propagating down a channel under the influence of Coriolis
force.  In essence, $u$ in the equation can be regarded as the
fluid velocity in the $x-$direction. The physical parameter
$\gamma $ measures the effect of the Earth's rotation. More
details about the Ostrovsky equation can be found e.g. in
\cite{O78,CIL07,LY07,St06}.

In the perturbation theory $\gamma\ll 1$ plays the role of a small
parameter. In order to ensure that the perturbation is decaying
fast enough at $x=\pm \infty$ we take the one-soliton KdV solution
in the form \b  u_s = 2\kappa_1^2/\sinh^2 z  \e which  can be
obtained formally from (\ref{1s}) for $\kappa_1 \xi_0 = \pi i /2$.
It is not continuous at $z=0$ but decays fast enough at $x=\pm
\infty$.  Using the fact that \b \frac{\mathrm{d}}{\mathrm{d}
z}\coth z = -\frac{1}{\sinh^2 z}+2\delta(z) = -\frac{1}{\sinh^2
z}+\frac{\mathrm{d}}{\mathrm{d} z}[\theta(z)-\theta(-z)]  \e
\cite{delta} we obtain \b P[u_s]=\gamma
\partial^{-1} u_s =-2\gamma \kappa_1 [\coth z-\theta(z)+\theta(-z)], \e  which is
an odd function of $z$ and then (\ref{kapa t}) gives
$\kappa_{1,t}=0$. Thus the amplitude of the soliton does not
change under this perturbation. The computation of (\ref{xi t})
gives a correction to the velocity of the soliton: \b
\xi_t=4\kappa_1^2+\frac{\pi^2\gamma}{8 \kappa_1^2}. \nonumber \e

\section{Conservation laws and perturbed soliton equations}\label{sec:4}

It is well known \cite{ZF71,ZMNP} that the KdV equation is a
completely integrable Hamiltonian system and possesses
infinitely-many integrals of motion. These integrals can be
constructed from the scattering coefficients $a(k)$ of the
associated spectral problem (\ref{SPKdV}) and are polynomials of
the dependent variable $u(x,t)$ and its $x$-derivatives:
\begin{eqnarray}\label{eq:4.1}
I_n=\int_{-\infty}^{\infty}P_n(u,u_x,u_{xx}, \dots)\, {\rm d}x,
\end{eqnarray}
where $P_n$ is a polynomial with respect to $u$ and its
derivatives. Since  $a(k)$ is time-independent, it can be viewed
as generating functional of integrals of motion $a_k$ \cite{ZMNP}:
\begin{eqnarray}\label{eq:4.2}
\mbox{ln}\, a(k)=\sum_{s=0}^{\infty}{I_{s+1}\over (2{\rm i}k)^s}.
\end{eqnarray}
Skipping the details (see e.g. \cite{ZMNP}), we provide here only
the list of the first few integrals of motion:
\begin{eqnarray}\label{eq:4.3a}
I_1&=&-{1\over 2}\int_{-\infty}^{\infty}u(x)\,{\rm d}x;\\
\label{eq:4.3b}
I_2&=&-{1\over 2}\int_{-\infty}^{\infty}u(x)^2\,{\rm d}x;\\
\label{eq:4.3c}
I_3&=&-{1\over 2}\int_{-\infty}^{\infty}\left(u_x^2(x)+2u^3(x)\right)\,{\rm d}x;\\
\label{eq:4.3d} I_4&=&-{1\over
2}\int_{-\infty}^{\infty}\left(u_{xx}^2-5u^2u_{xx}+5u^4\right)\,{\rm
d}x;
\end{eqnarray}
The KdV equation (\ref{eq:KdV}) can be written as a Hamiltonian
system
\begin{equation}\label{eqH1}
 u_{t}=\frac{\partial}{\partial x}\frac{\delta H}{\delta u(x)},
\end{equation}
where the symbol $\delta/\delta u$ denotes variational derivative.
Moreover, (\ref{eqH1}) can be further represented in its
Hamiltonian form with a Hamiltonian $H$:
\begin{equation}\label{eqH2}
 u_{t}=\{u, H\}.
\end{equation}
where $H=I_3$ (\ref{eq:4.3c}). The Poisson bracket is defined as
\begin{equation}\label{eqH3}
 \{F, G\}\equiv \int\frac{\delta F}{\delta u(x)}\frac{\partial}{\partial x}\frac{\delta G}{\delta
 u(x)}\text{d}x.
\end{equation}
The first three integrals of motion
(\ref{eq:4.3a})--(\ref{eq:4.3c}) have the same interpretation for
all members of KdV hirarchy: The first one, $I_1$ is related to
the algebraic structure  of the Poisson bracket (\ref{eqH3}): it
follows from the presence of the operator $\partial/\partial x$ in
the Poisson brackets. The integral (\ref{eq:4.3b}) has a meaning
of a momentum. It is related to the translation invariance of the
Hamiltonian. Since $H[u(x+\varepsilon)]-H[u(x)]\equiv 0$, the
expansion of $\int(H[u(x+\varepsilon)]-H[u(x)])\text{d}x$  in
$\varepsilon$ about $\varepsilon=0$ gives (note that $u(x)=\delta
P/\delta u$)
\begin{equation}
0=\int\frac{\delta H}{\delta u(x)}\frac{\partial u}{\partial
x}\text{d}x=\int\frac{\delta H}{\delta u(x)} \frac{\partial
}{\partial x}\frac{\delta P}{\delta u(x)}\text{d}x\equiv \{P,
H\}=P_t,\nonumber
\end{equation}
Consider now the perturbed KdV equation (\ref{PKdVH}).

We will seek the integrals of motion for the perturbed equation
$\tilde{I}_k$ in the form $\tilde{I}_k= I_k + \Delta I_k$,
$k=1,2,\dots$. Here $\Delta I_k$ can be viewed as a correction to
the integrals of motion of the unperturbed equation (\ref{eq:KdV})
coming from the perturbations $P[u]$. After a direct integration
of (\ref{PKdVH}),  one gets:
\begin{eqnarray}\label{eq:4.4}
\Delta I_1 = \int_{-\infty}^{\infty} P[u] \, {\rm d}x.
\end{eqnarray}
Then, multiplying (\ref{eq:KdV}) by $u(x,t)$, and integrating
leads to:
\begin{eqnarray}\label{eq:4.5}
\Delta I_2 = 2\int_{-\infty}^{\infty} u P[u] \, {\rm d}x,
\end{eqnarray}
and so on.

As an illustrative example we will take again the Ostrovsky
equation (\ref{ostr}). Due to the concrete choice of the
perturbation in the right hand side of (\ref{ostr}), the integrals
in (\ref{eq:4.4}) and (\ref{eq:4.5}) vanish, so the perturbations
do not contribute to these integrals: the first two integrals of
motion are the same as for the KdV equation. The nontrivial
contributions of perturbations to the integrals of motion in the
Hamiltonian $I_3$ are:
\begin{eqnarray}\label{eq:4.6}
\Delta I_3 = {\gamma \over 2}\int_{-\infty}^{\infty} (\partial
^{-1}u)^2 \, {\rm d}x.
\end{eqnarray}
Note also, that there is no second Hamiltonian formulation for the
Ostrovsky equation, compatible with the one given above, i.e. the
equation is not bi-Hamiltonian -- indeed (\ref{ostr}) is not
completely integrable for $\gamma\neq 0$, \cite{CIL07}.

\section{Perturbations to the equations of the Camassa-Holm hierarchy}\label{sec:5}

Closely related to the KdV hierarchy is the hierarchy of the
Camassa-Holm (CH) equation \cite{CH93}. This equation has the form
\begin{equation}\label{eq1}
 u_{t}-u_{xxt}+2\omega u_{x}+3uu_{x}-2u_{x}u_{xx}-uu_{xxx}=0,
\end{equation}
where $\omega$ is a real constant. It is integrable with a Lax
pair \cite{CH93}

\b \label{eq3} \Psi_{xx}&=&\Big(\frac{1}{4}+\lambda
(m+\omega)\Big)\Psi
 \\\label{eq4}
\Psi_{t}&=&\Big(\frac{1}{2\lambda}-u\Big)\Psi_{x}+\frac{u_{x}}{2}\Psi+\gamma\Psi
\e where $m\equiv u-u_{xx}$ and $\gamma$ is an arbitrary constant.

Both CH and KdV equations appeared initially as models of the
propagation of two-dimensional shallow water waves over a flat
bottom. More about the physical relevance of the CH equation can
be found e.g. in
\cite{CH93,J02,J03,DGH03,DGH04,I07,CL09,IK07,IK07a}.  The paper
\cite{Lak07} suggests that KdV and CH might be relevant to the
modelling of tsunami waves (see also the discussion in
\cite{CJ08}).

While all smooth data yield solutions of the KdV equation existing
for all times, certain smooth initial data for CH lead to global
solutions and others to breaking waves: the solution remains
bounded but its slope becomes unbounded in finite time (see
\cite{CE98, C00, BC07}).  The solitary waves of KdV are smooth
solitons, while the solitary waves of CH, which are also solitons,
are smooth if $\omega > 0$ \cite{CH93,J03} and peaked (called
``peakons" and representing weak solutions) if $\omega=0$
\cite{CH93,CE98-2,BBS98,CM00,L05-2}. Both solitary wave forms for
CH are stable \cite{CS00, CM01, CS02}.

It could be pointed out that the peakons appear also as travelling
wave solutions of greatest height (for the governing equations for
water waves), cf. \cite{C06,CE07,T96}.

In geometric context, the CH equation arises as a geodesic
equation on the diffeomorphism group (if $\omega=0$) \cite{C00,
CK03, CK06, K04} and on the Bott-Virasoro group (if $\omega > 0$)
\cite{M98}.

CH equation also allows for solutions with compactly supported
$m(x,t)$, \cite{C05}, however $u(x,t)$ looses instantly its
compact support, whether $\omega \ne 0$ \cite{H06} or $\omega = 0$
\cite{L07}.

The problem of perturbation of the CH equation arises when one
deals with model equations that are in general non-integrable but
close to the CH equation. A perturbation could appear for example
when one takes into account the viscosity effect \cite{SS07}.
Another possible scenario comes from the so-called '$b$-equation'
\cite{DHH,HW} \b m_t+b\omega u_x+bmu_x+m_xu=0. \nonumber \e The
$b$-equation generalizes the CH equation and is integrable only
for $b=2$ (when it coincides with the CH)  and $b=3$ (then known
as Degasperis-Procesi equation) \cite{MN,HW,I05}. Qualitatively,
the DP equation exhibits most of the features of the CH equation,
e.g. the infinite propagation speed for DP was established in
\cite{H05}. In \cite{Lak07} it is suggested that DP (as well as
CH) might be relevant to the modelling of tsunami waves (see also
the discussion in \cite{CJ08}).

The hydrodynamic relevance of the $b$-equation is discussed e.g.
in \cite{J03a,I07}. Therefore, the solutions of the $b$-equation
for values of $b$ close to $b=2$ can be analyzed in the framework
of the CH-perturbation theory. We can represent the equation as a
CH perturbation \b m_t+2\omega u_x+2mu_x+m_xu=(2-b)(\omega
u_x+mu_x)\equiv P[u], \nonumber \e

\n for a small parameter $\epsilon=b-2$.

\subsection{Inverse scattering and generalised Fourier transform for the CH spectral problem}

The CH spectral problem (\ref{eq3}) can be handled in a way,
similar to the one, already outlined. For simplicity we consider
the case where $m$ is a Schwartz class function, $\omega
>0$ and $m(x,0)+\omega > 0$. Let is introduce the notation $q(x,t)=m(x,t)+\omega$. Then one can show that
$q(x,t) > 0$ for all $t$ \cite{C01}.  Let
$k^{2}=-\frac{1}{4}-\lambda \omega$, i.e. \b \label{lambda}
\lambda(k)= -\frac{1}{\omega}\Big( k^{2}+\frac{1}{4}\Big).\e The
spectrum of the problem (\ref{eq3}) under these conditions is
described in \cite{C01}. The continuous spectrum in terms of $k$
corresponds to $k$ -- real. The discrete spectrum consists of
finitely many points $k_{n}=i\kappa _{n}$, $n=1,\ldots,N$ where
$\kappa_{n}$ is real and $0<\kappa_{n}<1/2$. The continuous
spectrum vanishes for $\omega = 0$, \cite{CM99}.

All results (\ref{eq6}) -- (\ref{eq206}) remain formally the same
with the exception of (\ref{dr}) which now has the form
\cite{CI06,CGI,CGI07}  \b a(k)=\exp \left(-i\alpha k
-\frac{1}{2\pi i}
\int_{-\infty}^{\infty}\frac{\ln(1-|\mathcal{R}^{\pm}(k')|^2)}{k'-k}\mathrm{d}k'
\right)\prod_{j=1}^{N}\frac{k-i\kappa_j}{k+i\kappa_j}.\label{drCH}\e
\n where \b \alpha &=& \int
_{-\infty}^{\infty}\Big(\sqrt{\frac{q(x)}
{\omega}}-1\Big)\text{d}x \nonumber \\
&=& \sum_{n=1}^{N}\ln\Big(\frac{1+2\kappa_n}{1-2\kappa_n} \Big)^2+
\frac{4}{\pi }\int _{0}^{\infty}\frac{\ln ( 1 - |\mathcal{R}^{+}(
\widetilde{k})|^2)}
 {4\widetilde{k}^2+1}\text{d}\widetilde{k} \nonumber \e
\n is one of the CH integrals of motion (Casimir).

With the asymptotics of the Jost solutions and (\ref{eq3}) one can
show that \begin{equation}\label{eq23aa}
L_{\pm}F^{\pm}(x,k)=\frac{1}{\lambda}F^{\pm}(x,k)\qquad
L_{\pm}F^{\pm}_n(x)=\frac{1}{\lambda_n}F^{\pm}_n(x),
\end{equation}
\n where $\lambda_n=\lambda(i\kappa_n)$; $F^{\pm}$ are again the
squares of the Jost solutions as in (\ref{eq23}) and  \b
L_{\pm}=(\partial^2-1)^{-1}\Big[4q(x)-2\int_{\pm
\infty}^{x}\text{d}y \, m'(y)\Big]\cdot\label{eq44}\e
\n is the recursion operator. The inverse of this operator is also
well defined.

The completeness relation for the eigenfunctions of the recursion
operator is \cite{CGI07}
\b \frac{\omega}{\sqrt{q(x)q(y)}} \theta(x-y)=-\frac{1}{2\pi i}
\int_{-\infty}^{\infty}\frac{F^-(x,k)F^+(y,k)}{ka^2(k)}\text{d}k
\phantom{**********}\nonumber\\ + \sum_{n=1}^{N}\frac{1}{i\kappa_n
\dot{a}_n^2}\Big[\dot{F}_n^-(x)F_n^+(y)+F_n^-(x)\dot{F}_n^+(y)-
\Big(\frac{1}{i\kappa_n}+\frac{\ddot{a}_n}{\dot{a}_n}\Big)
F_n^-(x)F_n^+(y)\Big].\label{eq24}\e Therefore $F^{\pm}$,
$F^{\pm}_n$ and $\dot{F}^{\pm}_n$ can be considered as
'generalised' exponents. Like in the KdV case it is possible to
expand $m(x)$ and its variation over the aforementioned basis, or
rather the quantities that are determined by $m(x)$ and $\delta
m(x)$, \cite{CGI07}: \begin{equation}\label{eq36}
\omega\Big(\sqrt{\frac{\omega}{q(x)}}-1\Big)= \pm\frac{1}{2\pi
i}\int_{-\infty}^{\infty}\frac{2k \mathcal{R}^{\pm}(k)}
{\lambda(k)} F^{\pm}(x,k)\text{d}k +
\sum_{n=1}^{N}\frac{2\kappa_n}{\lambda_n}R_n^{\pm}F_n^{\pm}(x);
\end{equation}
\b \frac{\omega}{\sqrt{q(x)}}\int_{\pm \infty}^{x}  &\delta
\sqrt{q(y)}&\text{d}y = \frac{1}{2\pi
i}\int_{-\infty}^{\infty}\frac{i}{\lambda(k)}\delta
\mathcal{R}^{\pm}(k)F^{\pm}(x,k)\text{d}k \nonumber  \\
& \pm & \!\! \sum_{n=1}^{N}\Big[\frac{1}{\lambda_n}(\delta
R_n^{\pm}-R_n^{\pm}\delta \lambda_n )F_n^{\pm}(x)
+\frac{R_n^{\pm}}{i\lambda_n}\delta \kappa_n \dot{F}_n^{\pm}(x)
\Big] \label{eq43a} \e
The expansion coefficients as expected are given by the scattering
data and their variations. This makes evident the interpretation
of the ISM as a generalized Fourier transform. Now it is
straightforward to describe the hierarchy of Camassa-Holm
equations. To every choice of the function $\Omega(z)$, known also
as the dispersion law we can put into correspondence the nonlinear
evolution equation (NLEE) that belongs to the Camassa-Holm hierarchy:
\begin{equation}\label{eq:CH}
\frac{2}{\sqrt{q}}\int_{\pm \infty}^{x}(\sqrt{q})_t \text{d}y
+\Omega(L_{\pm})\Big(\sqrt{\frac{\omega}{q}}-1\Big)=0.
\end{equation}
\n An equivalent form of the equation is \b
q_t+2q\tilde{u}_x+q_x\tilde{u}=0,\qquad
\tilde{u}=\frac{1}{2}\Omega(L_{\pm})\Big(\sqrt{\frac{\omega}{q}}-1\Big).\label{eq47}
\e The choice $\Omega(z)=z$ leads to $\tilde{u}=u$ and thus to the
CH equation (\ref{eq1}). Other choices of the dispersion law and
the corresponding equations of the Camassa-Holm hierarchy are
discussed in \cite{CGI07,I07DCDS}.

By virtue of the expansions (\ref{eq36}) and (\ref{eq43a}) the
NLEE (\ref{eq:CH}) is equivalent to the following linear evolution
equations for the scattering data: \b \mathcal{R}_t^{\pm}(k)\mp i
k \Omega(\lambda^{-1})
\mathcal{R}^{\pm}(k)=0, \label{eq48}\\
R_{n,t}^{\pm}\pm \kappa_n \Omega(\lambda_n^{-1})R_n^{\pm}=0,\label{eq49} \\
\kappa_{n,t}=0.\label{eq50}\e
The time-evolution of the scattering data for the CH equation
(\ref{eq1}) can be computed from the above formulae for
$\Omega(z)=z$, see also \cite{CI06,CGI}.

\subsection{Perturbation theory for the CH hierarchy}

Let us start with a perturbed equation of the CH hierarchy of the
form \b q_t+2q\tilde{u}_x+q_x\tilde{u}=P[u], \qquad
\tilde{u}=\frac{1}{2}\Omega(L_{\pm})\Big(\sqrt{\frac{\omega}{q}}-1\Big),\label{PCH}
\e
\n where again, $P[u]$ is a small perturbation, by assumption in
the Schwartz-class. It is useful to write (\ref{PCH}) in the form
\begin{equation}\label{eq:PCH}
\frac{2}{\sqrt{q}}\int_{\pm \infty}^{x}(\sqrt{q})_t \text{d}y
+\Omega(L_{\pm})\Big(\sqrt{\frac{\omega}{q}}-1\Big)=\frac{1}{\sqrt{q}}\int_{\pm
\infty}^{x}\frac{P(y)}{\sqrt{q(y)}} \text{d}y.
\end{equation}
With the completeness relation (\ref{eq24}) one can deduce the
gereralised Fourier expansion for expressions, like the one on the
right-hand side of (\ref{PCH})
\begin{theorem} \label{GFE-CH} Assuming that $f^+$ and $f^-$ are not linearly dependent at $x=0$ and $g(x)\in
\mathcal{S}(\mathbb{R})$, the following expansion formulas hold:
\begin{eqnarray}
\frac{\omega}{\sqrt{q}}\int_{\pm
\infty}^{x}\frac{g(y)}{\sqrt{q(y)}} \text{d}y=&\pm &\frac{1}{2\pi
i}\int_{-\infty}^{\infty}\tilde{g}^{\pm}(k)F^{\pm}_{x}(x,k)\mathrm{d}k
\nonumber\\
&\mp&\sum_{j=1}^{N}\left(g^{\pm}_{1,j}\dot{F}^{\pm}_{j,x}(x)+
g^{\pm}_{2,j}F^{\pm}_{j,x}(x) \right),
 \label{expP}
 \end{eqnarray}
\n where $\dot{F}^{\pm}_{j}(x)\equiv [\frac{\partial}{\partial k}
F^{\pm} (x,k)]_{k=k_j}$ and the Fourier coefficients are \b
\tilde{g}^{\pm}(k)&=&\frac{1}{k a^2(k)}\left(g,F^{\mp} \right),
\quad  \nonumber \\
g^{\pm}_{1,j}&=&\frac{1}{k_j \dot{a}_j^2}\left(g,F^{\mp}_j
\right),\nonumber \\
g^{\pm}_{2,j}&=&\frac{1}{k_j
\dot{a}_j^2}\left[\left(g,\dot{F}^{\mp}_j\right)-\left(\frac{1}{k_j}+\frac{\ddot{a}_j}{\dot{a}_j}\right)\left(g,F^{\mp}_j\right)
\right].\nonumber\e
\end{theorem}

The substitution of the expansions (\ref{expP}) for $P[u]$,
(\ref{eq36}) and (\ref{eq43a}) into the perturbed equation
(\ref{eq:PCH}) gives the following expressions for the modified
scattering data:
\b \mathcal{R}_t^{\pm}\mp
ik\Omega(1/\lambda)\mathcal{R}^{\pm}&=&\mp \frac{i\lambda (P,F^{\mp})}{2ka^2(k)},\label{CHPScatData 1} \\
k_{j,t}&=&\frac{\lambda_j(P,F_j^{\mp})}{2k_j\dot{a}_j^2R_j^{\pm}}
\label{CHPScatData 3} \\
R_{j,t}^{\pm}-R_j^{\pm}\lambda_{j,t}&\pm&\kappa_j
\Omega(1/\lambda_j)R_j^{\pm}\nonumber
\\ &=&-\frac{\lambda_j}{2k_j\dot{a}_j^2}\left[(P,\dot{F}_j^{\mp})-
\left(\frac{1}{k_j}+\frac{\ddot{a}_j}{\dot{a}_j}\right)\left(P,F^{\mp}_j\right)\right],
\label{CHPScatData 2} \e \n From (\ref{CHPScatData 2}) we obtain
the following for the coefficient $b_j$: \b
b_{j,t}+\kappa_j\Omega(1/\lambda_j)b_j=-\frac{\lambda_j}{4\kappa_j
\dot{a}_j} \left(P, b_j^2\dot{F}_j^+ - \dot{F}_j^-\right).
\nonumber \e The 'perturbed' solution for the hierarchy in the
adiabatic approximation can be recovered from the following
expansion for $\tilde{u}(x)$ with the 'modified' scattering data
keeping the unperturbed 'generalised' exponents: \b \tilde{u}(x) =
\pm \frac{1}{2\pi i}\int_{-\infty}^{\infty}\frac{k
\Omega(1/\lambda(k))}{\omega
\lambda(k)}\mathcal{R}^{\pm}(k)F^{\pm}(x,k)\text{d}k
+\sum_{n=1}^{N}\frac{\kappa_n\Omega(1/\lambda_n)}{\omega\lambda_n}R_n^{\pm}F_n^{\pm}
(x).\nonumber \e This formula follows from the second part of
(\ref{eq47}) and (\ref{eq36}). Note that for the CH equation
(\ref{eq1}) $\tilde{u}\equiv u$.

\section{Discussions }\label{sec:6}

We have presented a review of some aspects of the perturbation
theory for integrable equations using as main examples the KdV and
CH hierarchies.

In our derivations we used completeness relations that are valid
only given the assumption that the Jost solutions $f^+$ and $f^-$
are linearly independent at $x=0$. The case when this condition is
not satisfied is quite exceptional, however this is exactly the
case when one has purely soliton solution \cite{E81,IKK94}. Then
one has to take into account a nontrivial contribution from the
scattering data at $k=0$ \cite{He90c} and some of the presented
results require modification. E.g. (\ref{xi t}) should be
\cite{karp1979b} \b
\xi_t=4\kappa_1^2-\frac{1}{4\kappa_1^3}\int_{-\infty}^{\infty}
P[u_s(z)](z\mathrm{sech}^2 z +\tanh z+\tanh^2 z)\mathrm{d} z.
\label{xi t'} \e \n In the presented example with the Ostrovsky
equation \b \int_{-\infty}^{\infty} P[u_s(z)] \tanh^2 z \mathrm{d}
z=0 \label{shelf}\e since $P(z)$ is an odd function and the
additional term does not contribute. The meaning of the condition
(\ref{shelf}) is that no shelf is formed behind the soliton
\cite{karp1979b,He90c}. The presence of shelf for KdV equation is
observed e.g. under the perturbation $P[u]=\epsilon u$
\cite{karp1979b,L80}.

The corrections to the conservation laws due to perturbations have
been used in studies of the effects of the disturbance on the
initial soliton \cite{karp1979b,kaup1977}, or as a correctness
check of results obtained otherwise \cite{kn1978a}.

The evaluation of the perturbation terms for the CH hierarchy
could be technically difficult due to the complicated form of the
CH multisoliton solutions \cite{C01,CGI}. However the limit
$\omega \to 0$ leads to the relatively simple peakon solutions.
Therefore, using the presented general formulae one should be able
to access the perturbations of the peakon parameters.

\subsection*{Acknowledgements}
The authors are grateful to Prof. A.
Constantin and Prof. V. S. Gerdjikov for numerous useful
discussions, and to an anonymous referee for useful suggestions. The present work is partially supported by INTAS grant
No 05-1000008-7883.

\end{document}